\documentclass[twocolumn,aps,prb,floatfix,superscriptaddress]{revtex4-2}
\usepackage[utf8]{inputenc}
\usepackage{graphicx}
\usepackage{multirow}
\usepackage{float}
\usepackage{hyperref}
\usepackage{color}
\usepackage{amsmath}

\newcommand{\vect}[1]{{\ensuremath{\boldsymbol{{#1}}}}}
\newcommand{\vk}{\vect{k}}
\newcommand{\vq}{\vect{q}}
\begin{document}

\title{Topological excitons in moiré MoTe$_2$/WSe$_2$ heterobilayers}

\author{Paul Froese}
\affiliation{Department of Physics, University of Toronto, Ontario M5S 1A7, Canada}
\author{Titus Neupert}
\affiliation{Department of Physics, University of Zurich, Winterthurerstrasse 190, 8057 Zurich, Switzerland} 
\author{Glenn Wagner}
\affiliation{Department of Physics, University of Zurich, Winterthurerstrasse 190, 8057 Zurich, Switzerland}

\begin{abstract}
Due to the presence of flat Chern bands, moiré transition metal dichalcogenide (TMD) bilayers are a platform to realize strongly correlated topological phases of fermions such as fractional Chern insulators. TMDs are also known to host long-lived excitons, which inherit the topology of the underlying Chern bands. For the particular example of MoTe$_2$/WSe$_2$ heterobilayers we perform a time-dependent Hartree-Fock calculation to identify a regime in the phase diagram where the excitons themselves form a topological flat  band. This paves a way towards realizing strongly correlated states of bosons in moiré TMDs.
\end{abstract}

\maketitle

\section{Introduction}

\begin{figure}
    \centering
    \includegraphics[width=\columnwidth]{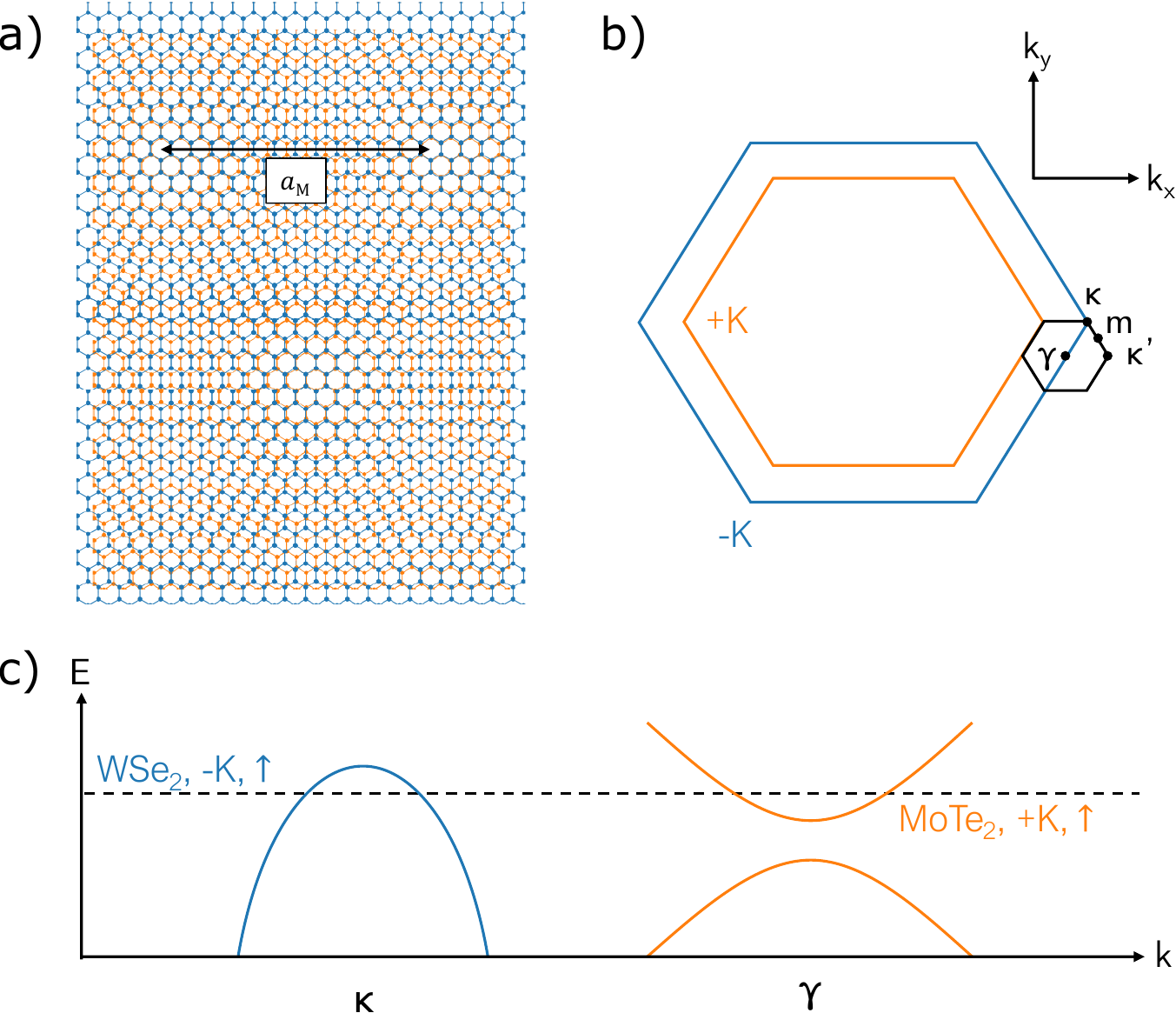}
    \caption{a) The lattice structure of an untwisted MoTe$_2$/WSe$_2$ heterobilayer. The unit cell of the moiré lattice with moiré lattice constant $a_\mathrm{M}$ contains $13 \times 13$ MoTe$_2$ unit cells (orange) and $14 \times 14$ WSe$_2$ unit cells (blue). b) The folding scheme of the moiré Brillouin zone (mBZ). Due to unequal lattice constants, the BZs of the bottom MoTe$_2$ layer (orange) and top WSe$_2$ layer (blue) have different sizes, leading to a folded mBZ (black). The $+K$ valley of the MoTe$_2$ layer and the $-K$ valley of the WSe$_2$ layer are labelled, as well as high symmetry points in the mBZ. 
    c) The Fermi surface at half-filling arises from an electron band at the $\gamma$ point and a hole band at the $\kappa$ point. We describe the system with a three band effective model, whose bands are schematically pictured.}
    \label{fig:SP}
\end{figure}

In recent years, moiré materials have emerged as  ideal systems for studying exotic phases of matter \cite{moire_review}. The translational symmetry breaking due to the spatial moiré pattern leads to folding in the Brillouin zone (BZ). This allows experiments to zoom in to a part of the bandstructure of the individual layers which has desirable properties.  In particular, for correlated and topological phases, properties such as flat bands and uniform Berry curvature are relevant. Moiré materials are very tunable. Besides the moiré pattern itself, which can be modified using the twist angle, they can be manipulated by gate-doping and displacement fields. Furthermore, many different types of van der Waals materials can be stacked and a variety of substrates can be used \cite{VDW}, leading to a whole zoo of possibilities for moiré heterostructures. 

The field started with the study of twisted bilayer graphene, which can realize correlated insulators, strange metals, and unconventional superconductivity \cite{TBG_review}. The band topology of twisted bilayer graphene is most apparent in its correlated insulator states, some of which exhibit the quantized anomalous Hall effect \cite{TBG_QAH,AHE_TBG}, which arises from integer filling of a Chern band. At fractional filling of a Chern band, one can obtain a fractional Chern insulator (FCI) \cite{neupert,sheng,regnault}. Such states have been observed at a finite magnetic field in twisted bilayer graphene~\cite{FCI_TBG}.

Moiré transition metal dichalcogenides (TMDs) offer a whole new set of possibilities. In the case of the TMDs, the moiré pattern can arise from twisted homobilayers. In particular, twisted MoTe$_2$ homobilayers have been recently shown to host zero field fractional Chern insulators \cite{cai2023signatures,zeng2023integer,Xu2023,Park2023}.  

Another class of moiré TMD materials are untwisted AB stacked TMD heterobilayers, where the moiré pattern arises from the different lattice constants of the constituent TMDs rather than from twisting. A material platform that has recently attracted interest is the AB stacked heterobilayer MoTe$_2$/WSe$_2$. Transport experiments have demonstrated the quantized anomalous Hall (QAH) effect \cite{MoTeWSe_QAH_exp1,MoTeWSe_Exp2}, which has motivated a number of theoretical works aiming to explain it \cite{MWth1,MWth2,MWth3,MWth4,MWth5,MWth6,MWth7,MWth8}. Recent magnetic circular dichroism (MCD) experiments established that the QAH is due to an intervalley coherent rather than a valley polarized state \cite{MCD}, requiring revisiting the theoretical models. A Hartree-Fock study indeed finds a topological intervalley coherent (IVC) state consistent with the observation of the QAH effect and the MCD measurements~\cite{xie2024topological}.

Given a system with topologically non-trivial bands, it is interesting to see how particle-hole excitations in those bands behave. Excitons have been studied in both twisted bilayer graphene \cite{Exciton_Band_Topology} and TMDs \cite{ExcitonsWTe2,mTMD1,mTMD2,qiu2024quantumgeometryprobedchiral}. In twisted bilayer graphene it has been shown using a time-dependent Hartree-Fock calculation that the excitons themselves can form relatively flat Chern bands \cite{Exciton_Band_Topology}. This motivated the proposal of excitonic fractional quantum Hall states \cite{Excitonic_FQHE}. At rational filling of the exciton bands, it is possible to obtain bosonic fractional quantum Hall states, making this a bosonic fractional Chern insulator.  While twisted bilayer graphene may not be the most promising platform for realizing these states, TMD heterobilayers are better-suited. This is because in heterobilayer TMDs, the two constituents of the exciton ---the electron and the hole--- are mostly supported in opposite layers, rendering the excitons very long lived. These excitons have already been studied experimentally \cite{gao2023excitonic,EX1,EX2,EX3,EX4}. They inherit the band topology of the underlying electron bands if the electronic band is a Chern band \cite{EBC1,EBC2,EBC3}.  The observation of the QAH effect in MoTe$_2$/WSe$_2$ heterobilayer motivates us to study that particular material. As a starting point for the theoretical modelling, we use the IVC state studied in Ref.~\onlinecite{xie2024topological}. Using Hartree-Fock and time-dependent Hartree-Fock, we then establish the exciton band topology and characterize its quantum geometry. 

\begin{figure*}
    \centering
    \includegraphics[width=\textwidth]{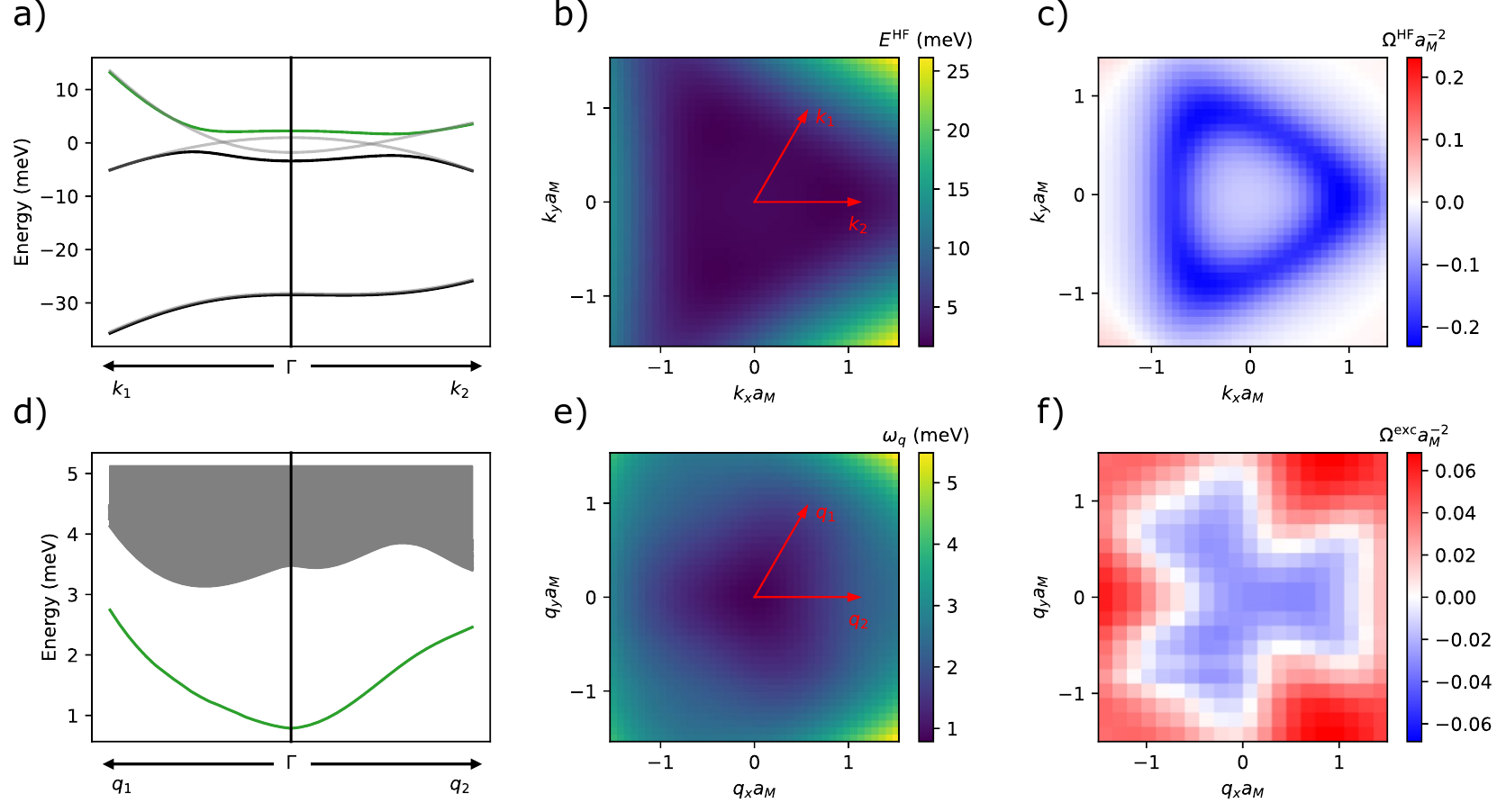}
    \caption{Top row: Hartree-Fock results. a) The quasiparticle energy spectrum along the $k_1$ and $k_2$ directions, which correspond to momentum cuts along $\Gamma - \kappa$ and $\Gamma - \kappa'$ respectively. The conduction band is highlighted in green, and the original non-interacting bands are shown in grey.  b) The conduction band. The $k_1$ and $k_2$ directions are shown. c) The Berry curvature of the conduction band. Bottom row: Time-dependent Hartree-Fock results. d) The exciton energy spectrum along the $q_1$ and $q_2$ directions. The lowest exciton band is separated from the continuum (grey) and highlighted in green. e) The lowest exciton band. The $q_1$ and $q_2$ directions are shown. f) The Berry curvature of the lowest exciton band. The input parameters for these results are $d_{sc}=10\ \mathrm{nm}, \delta E_\mathrm{D}=16\ \mathrm{meV}, 10/\epsilon_r\approx 0.4$.}
    \label{fig:Results}
\end{figure*}

\section{Methods}

The single-particle bandstructure of MoTe$_2$/WSe$_2$ heterobilayers can be described using the continuum model developed in Refs.~\onlinecite{Wu2019,moralesdurán2023magic}. At half-filling, there are two spin up bands crossing the Fermi level: A $+K$ band corresponding to the MoTe$_2$ layer, with an electron pocket at the $\gamma$ point of the moiré Brillouin zone, and a $-K$ WSe$_2$ band
with a hole pocket at the $\kappa$ point (see Fig.~\ref{fig:SP}). The spin polarization of these bands is a result of giant Ising spin-orbit coupling \cite{Ising_SOC1,Ising_SOC2,Ising_SOC3}. In order to recover a band alignment in agreement with MCD experiments \cite{MCD}, a Hund interaction is added to the continuum model. This yields an effective three band model \cite{xie2024topological}, whose bands are shown in Fig.~\ref{fig:SP}(b). The $-K$ WSe$_2$ band is well described by a simple parabolic dispersion $\xi_{-}(\vk)=-\frac{k^2}{2m}  + \delta E_\mathrm{D}$ with effective mass $m = 0.35m_{\mathrm{e}}$ and displacement field $\delta E_\mathrm{D}$. The $+K$ MoTe$_2$ band is not well separated and can be described with a $C_{3v}$ invariant massive Dirac Hamiltonian\begin{align}
    \hat{H}_{+}(\vk)
    =
    \begin{pmatrix}
        \varepsilon(\vk) + M(\vk) & -v_F(k_x +ik_y) + C(\vk) \\
        \textrm{h.c.} & \varepsilon(\vk) - M(\vk) \\
    \end{pmatrix}
    ,
\end{align}
where $\varepsilon(\vk) = A_0k^2$, $M(\vk) = M_0 + B_0k^2$ and $C(\vk)$ is a warping term given by
\begin{align}
    C(\vk) &= C_0[(k_x^2-k_y^2)+2ik_xk_y]
    .
\end{align}

The values for the parameters extracted from fitting the three-band model to the continuum model are $A_0 = 3903\ \mathrm{ meV \AA}^2, B_0 = -758\ \mathrm{ meV \AA}^2, C_0 = 6335\ \mathrm{ meV \AA}^2, M_0 = 13.22\ \mathrm{ meV}$, and  $v_{\mathrm{F}} = 448\ \mathrm{ meV \AA}$ \cite{xie2024topological}. Putting the contributions for both valleys together, we arrive at the effective three band Hamiltonian 
\begin{align}
    \hat{H}(\vk) = 
    \begin{pmatrix}
        \hat{H}_{+}(\vk) & 0 \\
        0 & \xi_{-}(\vk)
    \end{pmatrix} .
\end{align} 
The low energy model requires a momentum cutoff $K_\mathrm{c}$. In our calculations we take $K_\mathrm{c} a_\mathrm{M} = 1.5$, and consider a momentum range $|k_x|,|k_y| \leq K_\mathrm{c}$. The length scale $a_\mathrm{M}$ is the moiré lattice constant, $a_\mathrm{M} \approx 5\ \mathrm{nm}$.

We consider this noninteracting Hamiltonian together with the screened Coulomb interaction of the form 
\begin{equation}
    V(q) = \frac{e^2}{2\epsilon_0\epsilon_r\sqrt{q^2 + d_\mathrm{sc}^2}}
    ,
\end{equation}
where $d_\mathrm{sc}$ is the screening length and $\epsilon_r$ is the dielectric constant. The interactions hybridize the two bands near the Fermi level, leading to a gapped topological inter-valley coherent state \cite{xie2024topological}.

The new quasiparticle bands are found within a Hartree-Fock approximation \cite{xie2024topological}.  We neglect the Hartree term, as this only renormalizes the energy and does not play a role in the inter-valley coherent state \cite{xie2024topological}. The Fock order parameter is then solved for self-consistently.

% \begin{figure}
%     \centering
%     \includegraphics[width=\columnwidth]{figures/exciton_cartoon.pdf}
%     \caption{A particle is excited from the valence band to the conduction band. \GW{I think this can replace the current Fig. 1b}}
%     \label{fig:excitation}
% \end{figure}

Finally, we consider particle-hole excitations forming in the gapped Hartree-Fock bands. The resulting exciton energies and wavefunctions are calculated within a time-dependent Hartree-Fock approximation \cite{FL_theory}. This is done using an equation of motion method: First an exciton operator is constructed by taking a superposition of particle-hole excitations at all momenta $\vk$, and then its time dependence is found via a Heisenberg equation of motion. This amounts to calculating a commutator between the exciton operator and the Hamiltonian within the Hartree-Fock approximation \cite{Exciton_Band_Topology,TDHF}. Finally, Fourier transforming this equation of motion, we arrive at the following eigenvalue equation for the exciton energies $\omega_q$ and corresponding wavefunctions $\varphi_\vq(\vk)$, 
\begin{multline}
\label{eq:TDHF}
\omega_\vq \varphi_\vq(\vk) 
=
\left( E^\mathrm{HF}_c(\vk+\vq) - E^\mathrm{HF}_v(\vk) \right)\varphi_\vq(\vk)
\\
- 
\frac{1}{A}\sum_{\vk'}V_{cvvc}(\vk'+\vq,\vk,\vk-\vk')\varphi_\vq(\vk') 
.
\end{multline}
Here $E_c^\mathrm{HF}$ and $E_v^\mathrm{HF}$ are the energies of the Hartree-Fock conduction and valence bands. The interaction matrix element is 
\begin{equation}
    V_{\alpha\beta\gamma\delta}(\vk,\vk',\vq)
    :=
    V(q) \Lambda_{\alpha\delta}(\vk+\vq,\vk) \Lambda_{\beta\gamma}(\vk'-\vq,\vk')
\end{equation}
where the $\Lambda_{\alpha\beta}(\vk,\vk'):= \langle u_\alpha(\vk) | u_\beta(\vk') \rangle $ are form factors corresponding to the Hartree-Fock quasiparticle states $|u_\alpha(\vk)\rangle$. Equation (\ref{eq:TDHF}) can be solved numerically \cite{Kallin1984}.

To study the topology of the exciton states, we compute their Berry curvature~\cite{Fukui2005}. In general, the exciton band Berry curvature contains contributions from the underlying single-particle bands, the envelope function $\varphi_\vq(\vk)$, and the coupling between them~\cite{Exciton_Band_Topology,davenport2024interactioninducedcrystallinetopologyexcitons}. 

\section{Results}

The energies and topological properties of both the underlying quasiparticle bands and exciton bands are shown in Fig.~\ref{fig:Results}. Figure~\ref{fig:Results}(a-c) presents the Hartree-Fock results. We find that interactions indeed lead to a gapped insulating phase. The corresponding order parameter breaks translational symmetry due to the finite momentum pairing, which manifests as charge density wave behaviour in real space \cite{xie2024topological}. Furthermore, the quasiparticle bands are shown to be topologically non-trivial, with Berry curvature concentrated around the crossing point of the non-interacting bands. This phase inherits its topological nature from the massive Dirac Hamiltonian of the MoTe$_2$ layer. Specifically, this is a $p+ip$ IVC phase, where the IVC order parameter has a $k_x + ik_y$ momentum dependence, which was previously shown through a mean field analysis in the weak coupling limit, as well as a self-consistent Hartree-Fock calculation \cite{xie2024topological}. From the Wilson loop, which corresponds to the integral of the Berry curvature enclosed by the loop, the topology of the bands can be established. This is done by calculating square loops that enclose the origin in momentum space, and examining how the value of the Wilson loop changes as the radius of the loop is increased and more area is enclosed. Without interactions, the Wilson loop winds by a value approaching $-\pi$ for large radius, reflecting the massive Dirac cone at the origin in momentum space, which has a quantized winding of $-\pi$ at infinite radius. However, in the presence of interactions, it winds by $+\pi$, implying that the interactions have led to a change in the Chern number by $+1$.  Our Hartree-Fock calculation reproduces this result.

Figure \ref{fig:Results}(d-f) show the analogous plots for the exciton bands, calculated within the TDHF approximation. The lowest exciton band is well separated from the particle hole continuum, with a minimum at $\vq=0$. The band is relatively flat, with a bandwidth of roughly 1.5~meV in the range of momenta below the cutoff $K_\mathrm{c}$. This exciton band also has significant Berry curvature. 

\begin{figure}
    \centering
    \includegraphics[width=\columnwidth]{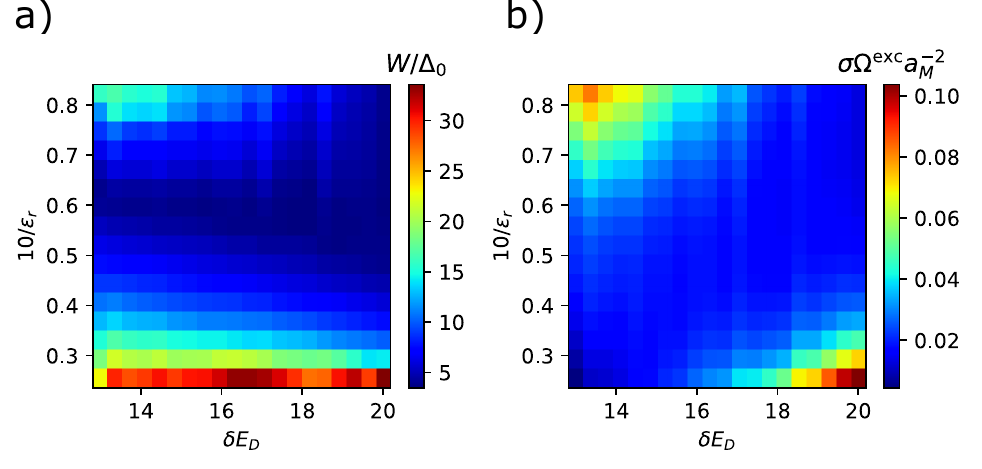}
    \caption{Dependence of the properties of the lowest exciton band on interaction strength and $\delta E_\mathrm{D}$ as tuning parameters. a) The bandwidth of the lowest exciton band divided by the minimum direct gap between the lowest two exciton bands. b) The standard deviation of the exciton Berry curvature. The system is most likely to host fractional Chern insulating states when both these quantities are small. For both diagrams $d_{sc} = 10\ \mathrm{nm}$.}
    \label{fig:phase_diagram}
\end{figure}

Though the necessary conditions are still debated, it is widely agreed that flat bands and uniform Berry curvature are desirable features when searching for fractional Chern insulators \cite{ledwith2022vortexability,IdealChern,MagicLine,BC1,BC2,BC3,BC4,BC5,BC6,BC7,BC8,BC9}. Varying the parameters of the model, we plot phase diagrams corresponding to these criteria. In particular we tune the interaction strength, characterized by $V \propto 10/\epsilon_r$, which can be controlled via the dielectric environment, such as the substrate. We also tune $\delta E_\mathrm{D}$, which determines the energy shift between the bands and can be understood as the result of a displacement field. In Fig.~\ref{fig:phase_diagram}(a) we plot $W/\Delta_0$, where $W$ is the bandwidth and $\Delta_0$ is the minimum direct gap between the lowest two exciton bands, and in Fig.~\ref{fig:phase_diagram}(b) we plot the standard deviation of the Berry curvature. In both phase diagrams, there is a region where both these values are minimal, occurring at $10/\epsilon_r \approx 0.6$ for small displacement field and widening as $\delta E_\mathrm{D}$ increases. This region is primarily constrained by the minimum of the bandwidth. These phase diagrams suggest that this region would be the most likely to host fractional Chern insulating states of excitons.

\section{Discussion}

\begin{figure}[t]
    \centering
    \vspace{5mm}
    \includegraphics[width=\columnwidth]{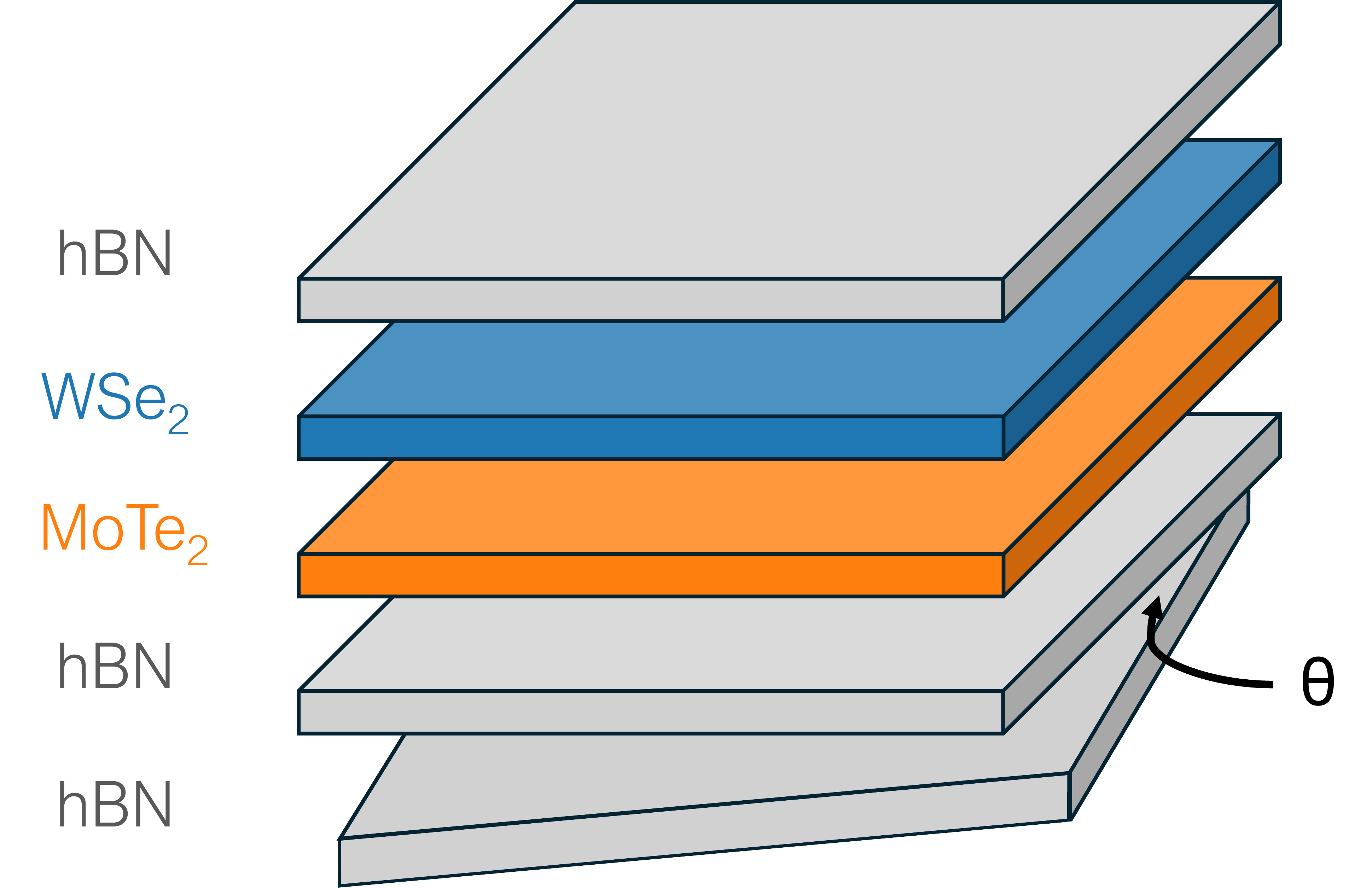}
    \caption{The proposed experimental set-up for realizing an FCI phase, with the MoTe$_2$/WSe$_2$ bilayer stacked between hBN layers. The bottom two hBN layers have a relative twist angle $\theta$ between them, creating a moiré potential which folds the exciton band to be flat and have uniform Berry curvature over the entire new smaller BZ.}
    \label{fig:stack}
\end{figure}

We have demonstrated that the AB-stacked TMD heterobilayer MoTe$_2$/WSe$_2$ hosts topological excitons. However, in order to realize an FCI phase, one requires a flat band throughout the entire Brillouin zone, whereas the exciton band we find is only flat in part of the moiré BZ defined by our cutoff $K_\mathrm{c}$. One way to obtain a flat band that fills the entire BZ is to apply a periodic potential which increases the lattice constant to be $\sim 1/K_\mathrm{c}$. This leads to a folding of the moiré BZ to an even smaller BZ. A natural candidate for inducing a periodic potential is the hBN substrate on which the TMDs lie. hBN has a large lattice constant mismatch relative to the TMDs and therefore there is no large-scale moiré pattern between the hBN and the TMD. However, by stacking two layers of hBN with a relative twist angle, one can engineer a large-scale periodic potential with a tunable moiré periodicity \cite{kim2023electrostatic,Woods2021,li2024moire}. If the TMD heterobilayer is stacked on top of this twisted hBN, it will experience a periodic potential. We therefore propose the following experimental set-up from  bottom to top: hBN-hBN-MoTe$_2$-WSe$_2$-hBN, where the two bottom hBN layers have a relative twist angle of $\theta$, as shown in Fig.~\ref{fig:stack}. 

It would be interesting to further study the strong correlation effects of these excitons to establish whether they can for example form a fractional quantum Hall state. A natural calculation would be to perform an exact diagonalization (ED) calculation on the exciton bandstructure calculated in our work. In this set up  the twist angle $\theta$ and the displacement field $\delta E_\mathrm{D}$ offer two useful tuning parameters.

Furthermore, it would be interesting to study the excitons in MoTe$_2$/WSe$_2$ experimentally. The rational filling of the exciton band can be achieved through optical pumping for example \cite{Huang2022,Mak2024,Regan2022}. Unlike the fermionic counterpart, which has a fractionally quantized Hall conductivity, the excitonic quantum Hall state is formed from electrically neutral excitons and therefore does not have a charge Hall response. Measuring a topological response of such a state -- for instance through thermal transport experiments -- is thus more challenging than for the charged counterparts. However, optical probes should show signatures of time-reversal symmetry breaking in such a state. In particular, Kerr rotation \cite{Tse2010} or MCD can be used as a probe, where the latter has already been applied to MoTe$_2$/WSe$_2$ bilayers, albeit in the absence of pumping \cite{MCD}. In order to probe the excitons, trion photoluminescence may be used, as it has been in twisted MoTe$_2$ homobilayers \cite{Cai2024}.

\section{Acknowledgements}

We acknowledge helpful discussions with Ajit Srivastava and with the authors of \cite{xie2024topological}. GW acknowledges funding from the University of Zurich postdoc grant FK-23-134. TN acknowledges support from the Swiss National Science Foundation through a Consolidator Grant (iTQC, TMCG-$2\_$213805). 

\bibliography{moirefti}

\setcounter{figure}{0}
\let\oldthefigure\thefigure
\renewcommand{\thefigure}{S\oldthefigure}

\setcounter{table}{0}
\renewcommand{\thetable}{S\arabic{table}}

\newpage
\clearpage

\begin{appendix}
\onecolumngrid
	\begin{center}\textbf{\large --- Supplementary Material ---}
    \end{center}

%\section{Continuum model}
%\PF{I would consider leaving this section out of the paper}

%\begin{align}
%   \mathcal{H}_\tau(\vect{r}) = 
%        \begin{pmatrix}
%            \mathcal{H}_\tau^b(\vect{r}) + \tau M_z & W_\tau (\vect{r}) \\
%            W_\tau^\dagger (\vect{r}) & \mathcal{H}_\tau^t(\vect{r}) - \tau M_z - \delta E_\mathrm{D}
%        \end{pmatrix} 
%\end{align}

%\begin{align}
%    \mathcal{H}_\tau^l(\vect{r}) = -\frac{\hat{\vect{p}}^2}{2m_l} + \tau \lambda_l(\hat{p}_x^3 - 3\hat{p}_x\hat{p}_y^2) + V_l(\vect{r})
%\end{align}

%\begin{align}
%    V_l(\vect{r}) = 2V_l\sum_{j=1,3,5} \cos(\vect{G}_j \cdot \vect{r} + \phi_l)
%\end{align}

%\begin{align}
%    W_\tau (\vect{r}) = W(1+\omega^\tau e^{-i\vect{G}_2 \cdot \vect{r}} + \omega^{2\tau} e^{-i\vect{G}_3 \cdot \vect{r}} )
%\end{align}

\section{Effective Hamiltonian}
As discussed in the main text, there are two relevant bands at the Fermi level. The isolated $-K$ WSe$_2$ band, with its minimum at the $\kappa$ point, is modelled with a parabolic dispersion
\begin{align}
    \xi_{-}(\vk) 
    =
    -\frac{\vk^2}{2m_t} - \mu + \delta E_\mathrm{D}
    ,
\end{align}
where $\mu$ is the chemical potential and $m_t = 0.35m_e$ is the effective mass of the top WSe$_2$ layer. The $+K$ MoTe$_2$ band at the $\gamma$ point is modelled by the massive Dirac Hamiltonian
\begin{align}
    \hat{H}_{+,\gamma}(\vk)
    =
    \begin{pmatrix}
        \varepsilon_0(\vk) + M(\vk) & -v_F(k_x +ik_y) + C_0(\vk) \\
        -v_F(k_x -ik_y) + C_0^*(\vk) & \varepsilon_0(\vk) - M(\vk) \\
    \end{pmatrix}
    ,
\end{align}
where
\begin{align}
    \varepsilon_0(\vk) &= A_0(k_x^2 +k_y^2) -\mu\\
    M(\vk) &= M_0 + B_0(k_x^2 +k_y^2)
\end{align}
and $C_0(\vk)$ is a warping term given by
\begin{align}
    C_0(\vk) &= C_0[(k_x^2-k_y^2)+2ik_xk_y]
    .
\end{align}

The values chosen for the parameters are $A_0 = 3903 \mathrm{ meV \AA}^2, B_0 = -758 \mathrm{ meV \AA}^2, C_0 = 6335 \mathrm{ meV \AA}^2, M_0 = 13.22 \mathrm{ meV}$, and  $v_F = 448 \mathrm{ meV \AA}$. Putting the contributions for both valleys together, we arrive at the effective three band Hamiltonian 
\begin{align}
    \hat{H}_0(\vk) = 
    \begin{pmatrix}
        \hat{H}_{+,\gamma}(\vk) & 0 \\
        0 & \xi_{-}(\vk)
    \end{pmatrix} 
    ,
\end{align}
written in the $\left\lbrace |\vk,\gamma, 1+\rangle, |\vk,\gamma, 2+\rangle, |\vk,\kappa, 3-\rangle  \right\rbrace$ basis. In order to find the energy bands, this Hamiltonian is diagonalized by a unitary matrix $U_0(\vk)$, i.e.
\begin{align}
    U_0(\vk)^\dagger \hat{H}_0(\vk) U_0(\vk) 
    = 
    \operatorname{diag}(\xi_{1+}(\vk),\xi_{2+}(\vk),\xi_{-}(\vk)), 
    \quad 
    \xi_{1+}(\vk) < \xi_{2+}(\vk).
\end{align} 

\section{Hartree-Fock approximation}
Let $c_\tau^\dagger(\vk)$ and $c_\tau(\vk)$ be electron creation and annihilation operators for the band basis. We restrict to the two bands of interest, i.e. the $\xi_-(\vk)$ and the $\xi_{2+}(\vk)$ bands. As now we only have one band from each valley, $\tau = \pm$ labels the valley, and we let $\xi_{+}(\vk) := \xi_{2+}(\vk)$. Introducing the interaction term, the full Hamiltonian is given by
\begin{align}
    \mathcal{H}
    = 
    \sum_{\vk} \sum_{\tau} \xi_\tau(\vk) c^\dagger_\tau(\vk) c_\tau(\vk)
    +
    \frac{1}{2A}\sum_{\vk\vk'\vq}\sum_{\tau \tau'} V^0(\vk,\vk',\vq)
    c^\dagger_\tau(\vk+\vq) c^\dagger_{\tau'}(\vk'-\vq) c_{\tau'}(\vk') c_\tau(\vk)
    .
\end{align}
Here $A$ is the area of the system. The matrix element $V^0(\vk,\vk',\vq)$ is defined as
\begin{align}
    V^0(\vk,\vk',\vq) := V(q)\Lambda^0_{\tau}(\vk+\vq,\vk)\Lambda^0_{\tau'}(\vk'-\vq,\vk')
\end{align}
with
\begin{align}
    \Lambda^0_{\tau}(\vk,\vk') :&= \langle u_0(\vk), \tau | u_0(\vk'), \tau \rangle \nonumber \\
    &= \sum_n [U_0(\vk)]_{n\tau}^*[U_0(\vk')]_{n\tau} 
    ,
\end{align}
where $| u_0(\vk), \tau \rangle$ are the eigenstates of $\hat{H}_0(\vk)$.
We do a standard Hartree-Fock approximation on the quartic interaction term, keeping only the Fock term. This approximation amounts to the following replacement:
\begin{align*}
  c^\dagger_\tau(\vk+\vq) c^\dagger_{\tau'}(\vk'-\vq) c_{\tau'}(\vk') c_\tau(\vk) \approx 
- \langle c^\dagger_\tau(\vk+\vq) c_{\tau'}(\vk') \rangle c^\dagger_{\tau'}(\vk'-\vq)  c_\tau(\vk)  
.
\end{align*}
Along with this replacement, we define the corresponding order parameter as
\begin{align}
    \label{eq:delta}
    \Delta_{\tau\tau'}(\vk) := -\frac{1}{2A}\sum_{\vk'}V^0(\vk,\vk',\vk'-\vk)(\rho_{\tau\tau'}(\vk') - \rho^\mathrm{eq}_{\tau\tau'}(\vk'))
    ,
\end{align}
where we have also introduced the density matrix $\rho_{\tau\tau'}(\vk) = \langle c^\dagger_\tau(\vk) c_{\tau'}(\vk) \rangle$, and a reference matrix $\rho^\mathrm{eq}_{\tau\tau'}(\vk)$, which is the non-interacting density matrix with $\mu$ at half-filling. The resulting mean-field Hamiltonian is
\begin{align}
    \mathcal{H}^{\text{MF}} = \sum_{\vk}\sum_{\tau\tau'} 
    \hat{H}_{\tau\tau'}(\vk)
    c_{\tau}^\dagger(\vk)c_{\tau'}(\vk) 
\end{align}
with
\begin{align}
    \hat{H}_{\tau\tau'}(\vk) = 
    \left(\delta_{\tau\tau'}\xi_\tau(\vk) + \Delta_{\tau'\tau}(\vk)\right)
\end{align}
We diagonalize the Hamiltonian with the unitary matrix $U(\vk)$ as follows
\begin{align}
    \label{eq:UHU}
    U^\dagger(\vk)\hat{H}(\vk)U(\vk) = \operatorname{diag}\left(E_v^\mathrm{HF}(\vk),E_c^\mathrm{HF}(\vk)\right)
    ,
\end{align}
where $E_v^\mathrm{HF}<E_c^\mathrm{HF}$ are the resulting quasiparticle valence and conduction bands respectively. The density matrix is calculated via
\begin{align}
    \rho_{\tau\tau'}(\vk) = \sum_n U^*_{\tau n}(\vk) U_{\tau n}(\vk) f\left( E_n^\mathrm{HF}(\vk) \right)
\end{align}
from which the order parameter can be recalculated and iterated self-consistently until convergence.

\section{Time-dependent Hartree-Fock approximation}
We introduce creation and annihilation operators for the HF bands, $d_\alpha^\dagger(\vk)$ and $d_\alpha(\vk)$, where $\alpha = c,v$ labels the conduction or valence band. The HF ground state is then defined as 
\begin{align}
    |\mathrm{GS} \rangle := \prod_{\vk} d_v^\dagger(\vk) |0\rangle
    .
\end{align}
The full Hamiltonian in the quasiparticle basis is
\begin{align}
\mathcal{H} = \sum_{\vk \alpha} E^\text{HF}_\alpha(\vk)d_\alpha^\dagger(\vk)d_\alpha(\vk) 
    +
    \frac{1}{2A}\sum_{\vk\vk'\vq}\sum_{\alpha\beta\gamma\delta}
    V_{\alpha\beta\gamma\delta}(\vk,\vk',\vq)
    d_\alpha^\dagger(\vk+\vq)
    d_\beta^\dagger(\vk'-\vq)
    d_\gamma(\vk')
    d_\delta(\vk)
\end{align}
where now the interaction matrix element is
\begin{align}
    V_{\alpha\beta\gamma\delta}(\vk,\vk',\vq)
    :=
    V(q) \Lambda_{\alpha\delta}(\vk+\vq,\vk) \Lambda_{\beta\gamma}(\vk'-\vq,\vk')
\end{align}
with form factors $\Lambda$ corresponding to the HF quasiparticle states $|u_\alpha(\vk)\rangle$, defined as
\begin{align}
    \Lambda_{\alpha\beta}(\vk,\vk')
    :=
    \langle u_\alpha(\vk) | u_\beta(\vk') \rangle
    .
\end{align}
We next introduce the neutral operator
\begin{align}
    b^\dagger(\vk,\vq) := d^\dagger_c(\vk+\vq)d_v(\vk)
    .
\end{align}
The time dependence of this operator follows the Heisenberg equation of motion:
\begin{align}
    \partial_t b^\dagger(\vk,\vq) = \frac{i}{\hbar} \left[\mathcal{H}, b^\dagger(\vk,\vq) \right]
    .
\end{align}
When calculating the interacting part of the commutator, we decouple any quartic terms in usual Hartree-Fock way, taking expectation values in the HF ground state, again keep only the Fock terms. This leads to the expression
\begin{align}
\left[\mathcal{H}, b^\dagger(\vk,\vq) \right] 
=
\left(E^\mathrm{HF}_c(\vk+\vq) - E^\mathrm{HF}_v(\vk)\right)b^\dagger(\vk,\vq)
-\frac{1}{A}\sum_{\vk'}V_{cvvc}(\vk+\vq,\vk',\vk'-\vk)b^\dagger(\vk',\vq)
.
\end{align}
In this derivation we have approximated the band indices $c$ and $v$ as conserved, since the excitons are long-lived. Next, we introduce the exciton operator
\begin{align}
\label{eq:exciton}
    \gamma^\dagger(\vq) := \sum_\vk \varphi_\vq(\vk)b^\dagger(\vk,\vq)
    .
\end{align}
Fourier transforming the equation of motion for $\gamma^\dagger(\vq)$, we find
\begin{align}
\omega_\vq\gamma^\dagger(\vq) 
=
\left[ \mathcal{H}, \gamma^\dagger(\vq) \right] 
=
\sum_k \varphi_\vq(\vk) \left[ \mathcal{H}, b^\dagger(\vk,\vq) \right] 
.
\end{align}
By expanding the $\gamma^\dagger(\vq)$ operator in the leftmost term and writing out the commutator in the rightmost term, we are left with the following eigenvalue equation for the exciton energy $\omega_\vq$ and wavefunction $\varphi_\vq(\vk)$:
\begin{align}
\omega_\vq \varphi_\vq(\vk) 
=
\left( E^\mathrm{HF}_c(\vk+\vq) - E^\mathrm{HF}_v(\vk) \right)\varphi_\vq(\vk)  
- 
\frac{1}{A}\sum_{\vk'}V_{cvvc}(\vk'+\vq,\vk,\vk-\vk')\varphi_\vq(\vk') 
.
\end{align}
We solve this equation by recasting it as a matrix equation,
\begin{align}
    \omega_\vq \varphi_\vq(\vk) \equiv \sum_{\vk'} \hat{T}_\vq(\vk;\vk')\varphi_\vq(\vk')
    ,
\end{align} and diagonalizing it numerically.

\section{Constructing exciton eigenstates}
The topology of the exciton band is encoded in the exciton eigenstates. In this section, we construct the relevant exciton eigenstates by closely following the procedure described in the supplemental material of \cite{Exciton_Band_Topology}. First, we introduce some notation for the eigenstates of our quasiparticle HF bands. The single particle eigenstates of the HF band can be written as
\begin{align}
    | \phi_{\vk,\alpha} \rangle = \sum_i u_{\vk,\alpha i} | \vk, i \rangle
    = \frac{1}{\sqrt{N}}\sum_{\vect{r} i} u_{\vk,\alpha i} e^{i\vk\cdot\vect{r}} | \vect{r}, i \rangle
    .
\end{align}
Here, $u_{\vk,\alpha i}$ are the wavefunctions found during the HF calculation, and can be extracted from the columns of the matrix $U(\vk)$ (c.f. \ref{eq:UHU}). $\alpha = c,v$ labels the HF band, and $i$ labels the states in the orbital basis.

We can then define the cell-periodic eigenstate as 
\begin{align}
    |u_{\vk,\alpha} \rangle :=
    e^{-i\vk\cdot\hat{\vect{r}}}| \phi_{\vk,\alpha} \rangle
    =
    \frac{1}{\sqrt{N}}\sum_{\vect{r} i} u_{\vk,\alpha i}| \vect{r}, i \rangle
\end{align}

We now construct a similar cell-periodic eigenstate for the excitons. Given the definition of the operator (\ref{eq:exciton}), the natural corresponding exciton state is
\begin{align}
\label{eq:exc2}
    |\text{exc},\vq \rangle := \sum_{\vk}\varphi_\vq(\vk)b^\dagger(\vk,\vq) | \mathrm{GS} \rangle
    =
    \sum_{\vk}\varphi_\vq(\vk)d_c^\dagger(\vk+\vq)d_v(\vk) | \mathrm{GS} \rangle
    .
\end{align}

However, we must make two important adjustments to this wavefunction in order to properly describe the exciton state. First, we do a particle-hole transformation in the valence band, so that we can work with two particle eigenstates. Secondly, we shift the momenta $(\vk+\vq, \vk)$ to $(\vk+\frac{\vq}{2}, \vk-\frac{\vq}{2})$. This is necessary to properly couple the centre of mass momentum of the exciton with its centre of mass position $\frac{\vect{r}_1 + \vect{r}_2}{2}$. Making these adjustments, the exciton eigenstate is defined as
\begin{align}
    |\psi^\text{exc}_\vq \rangle
    :=
    \sum_\vk \psi_\vq (\vk) |\phi_{\vk+\frac{\vq}{2},c}\rangle |\phi^*_{\vk-\frac{\vq}{2},v}\rangle
\end{align}
with $\psi_\vq(\vk) = \varphi_\vq(\vk - \frac{\vq}{2})$. Finally, putting these definitions together, the cell-periodic exciton state is
\begin{align}
    \label{eq:estate}
    |u_\vq^\text{exc} \rangle 
    :&=
    e^{-i\vq \cdot \left(\frac{\vect{r}_1 +\vect{r}_2}{2}\right)}|\psi^\text{exc}_\vq \rangle \nonumber \\
    &=  \sum_\vk \psi_\vq(\vk) \sum_{\vect{r}_1\vect{r}_2} \sum_{ij}
    u_{\vk+\frac{\vq}{2},ci}u^*_{\vk-\frac{\vq}{2},vj} 
    e^{i\vk\cdot(\vect{r}_1-\vect{r}_2)}
    |\vect{r}_1,i\rangle
    |\vect{r}_2,j\rangle
    .
\end{align}

\end{appendix}

\end{document}